\documentclass[useAMS,usenatbib,psfig,a4]{mn2e}
\usepackage{aas_macros}
\usepackage{graphicx}
\usepackage{epstopdf}

\title{Smooth Particle Lensing}
\author[Aubert, Amara \& Metcalf]{Dominique Aubert$^{1}$\thanks{dominique.aubert@cea.fr},  Adam Amara$^{1}$\thanks{aamara@cea.fr} \& R. Benton Metcalf$^{2}$\\
$^{1}$Service d'Astrophysique, CEA Saclay, Gif sur Yvette, 91191, France\\
$^{2}$Max Plank Institute for Astrophysics, Garching 8574, Germany}

\date{\today}

\begin{document}

\maketitle

\label{firstpage}

\begin{abstract}
We present a numerical technique to compute the gravitational
lensing induced by simulated haloes. It relies on a 2D-Tree domain
decomposition in the lens plane combined with a description of N-Body
particles as extended clouds with a non-singular density. This
technique is made fully adaptive by the use of a density-dependent
smoothing which allows one to probe the lensing properties of haloes
from the densest regions in the center or in substructures to the
low-density regions in the outskirts. 'Smooth Particle Lensing' has
some promising features. First, the  deflection potential, the
deflection angles, the convergence and the shear are direct and
separate end-products of the SPL calculation and can be computed at an
arbitrary distribution of points on the lens plane. Second, this
flexibility avoids the use of interpolation or a finite differentiation
procedure on a grid, does not require padding the region with zeros and focuses
the computing power on relevant regions.  The SPL algorithm is tested 
by populating isothermal spheres and ellipsoids with particles and then
comparing the lensing calculations to the classical FFT-based
technique and analytic solutions. We assess issues related to the
resolution of the lensing code and the limitations set by the simulations
themselves. We conclude by discussing how SPL can be used to predict
the impact of substructures on strong lensing and how it can be
generalized to weak lensing and cosmic shear simulations.  
\end{abstract}


\section{Introduction}
\label{intro}
The concordance model of cosmology has now been widely accepted as it is in agreement with an extensive range of observational probes \citep{2000ApJ...530...17W,2003astro.ph..5089V,1999Sci...284.1481B,2003ApJS..148..175S}. 
One of the central ingredient to the model is cold dark matter
(CDM). However, in order to draw detailed comparison between
theoretical models of CDM and the observed Universe high resolution
computational simulations are required. The simplest simulations
calculate the evolution of density perturbations in a Universe, where
the matter contribution is composed entirely of CDM particles
\citep{2003ApJS..145....1B,1998ApJ...499L...5M,1996ApJ...462..563N,1999AAS...194.5810E}.
The density distributions that these simulations produce cover a wide
range of scales, with an enormous wealth of information. The challenge
that we favirialce now is finding ways to efficiently mine this data and
compare it to observational probes. 

Compact collapsed objects, such as galaxies, groups and clusters, have
 been studied extensively in the simulations.  In doing so, a
 discrepancy known as 'the substructure problem' has been identified
 \citep{1999ApJ...524L..19M,1999ApJ...522...82K,2004astro.ph..5445W},
 where the number of small scale objects formed in simulated Milky
 Way's haloes is significantly greater than the number of observed satellites.
There is some ambiguity as to how dwarf galaxies are matched with
 subhalos of a given mass (see e.g. \citet{2002MNRAS.335L..84S},
 \citet{2004ApJ...608..663K}).  It is perhaps possible to resolve the
 problem for some mass range by adjusting this relation, but it is
 always the case that the simulations predict more subhalos than there
 are observed dwarf galaxies.  Whether this is a result of suppressed
 star formation in small halos or a deeper problem with the
 CDM model is not yet clear.

The flux ratios between images in strong gravitational lenses are powerful tools for studying 
substructure even when it is not associated with any observable stars or gas \citep{1998MNRAS.295..587M,2001ApJ...563....9M,2002ApJ...567L...5M,Dalal2002,2002ApJ...565...17C,MM02,cirpass2237}. A number of authors have investigated the implications of N-Body results on strong lensing.  One approach is to divide the study into two steps. The first step is to study the statistical properties of N-Body halo, such as the radial distribution of substructure. The second step is to predict the lensing implications separately by using analytic calculations \citep{2004ApJ...604L...5M}. This method has a number of drawbacks. For instance, they are dependent on the selection criteria used to identify substructure and analytic models for substructure lensing do not incorporate all the necessary effects.  Another approach has been to study the lensing properties of the simulations directly using ray-shooting. This approach also suffers from a number of drawbacks, which we will investigate here.  
 
 Both of these methods are limited by the resolutions of the N-Body simulations.  First, there is a mass resolution, set by the fact that halos are modeled using a finite number of particles. The current generation of simulations typically provide dark matter halos with 1$\rightarrow$10 million particles within the virial radius.  For a Milky Way-sized galaxy, this corresponds to mass resolution of $10^5\rightarrow10^6 M_\odot$. Any results below this mass scale can only be reached by extending what we see for larger mass scales \citep{2004ApJ...604L...5M}.  Numerical simulations are also limited in spatial length scales. The current generation of simulations have a spatial resolution of 50~$\rightarrow$~500~pc for a galaxy sized halo.  Although these scales result from an astonishing dynamic range for cosmological simulations, they still place major constraints on the strong lensing studies since strong lensing probes the very inner regions of galaxies, typically radii of 1-10  kpc, .
 
 Probing these inner regions in detail is also limited by the fact that the baryonic component of the galaxy plays an important role.  An ideal simulation would included dark matter as well as the gas, stars and their complex interactions.  Strong lensing studies of such simulations have been performed \citep{astro-ph/0306238}. However, it is well-known that the current generation of simulations that contain baryons struggle to reproduce realistic galaxies (see e.g. \citet{1995MNRAS.275...56N}, \citet{1997ApJ...478...13N}).  An alternative approach is to take the results of an N-Body dark matter simulation and add a model for a realistic galaxy and then allow the dark matter to relax in the new potential (\citet{2006MNRAS.tmp..276A}). Although somewhat ad hoc, this approach allows us to decouple the contributions due to observed baryonic distributions from those due to the dark matter in their halos.
 
\begin{figure} 
\centering
\resizebox{0.95\columnwidth}{0.675\columnwidth}{\includegraphics{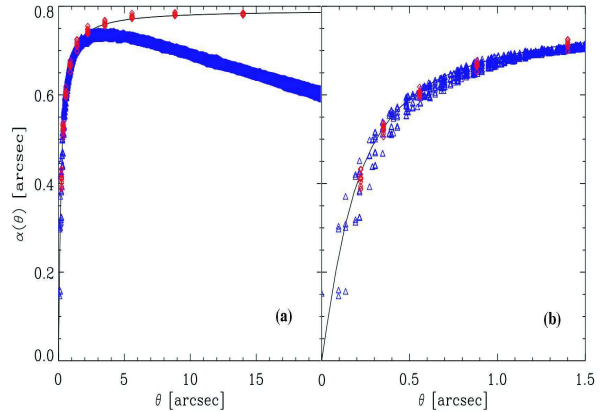}}
\caption{Comparison of the deflection angles calculated using the FFT method (blue triangles) and SPL method (red diamonds) with the analytic prediction (black curve).  On the left (panel (a)) we see that for large radii the FFT method becomes dominated by the repeating boundary conditions.  This is a problem that does not occur using the SPL method.  However, we know that the strong lensing region is very close to the centre. On the left (panel(b)) we see that the problem of repeatied boundary conditions is less pronounced.   Here a buffering of one quarter, i.e. one quarter of the FFT length is filled with zeros.}
\label{f:bound} 
\end{figure}

A final limitation of directly studying  the strong lensing of N-body halos is the technique used to perform the ray shooting. A popular method for calculating the lensing properties of object is to project it onto a 2D grid and to calculate the gravitational potential using an Fast Fourier Transform  (FFT) technique. This method has the advantage of being significantly faster to compute than others, scaling as NlogN and not N$^2$ as in the case of particle-particle calculations.  The FFT method, however, does suffer from two major drawbacks. The first is that the resolution of the lensing calculation is limited to grid size on which the FFT is performed, and the second is that using FFT routines introduces repeating boundary conditions into the problem. The latter means that the galaxy being studied is no longer an isolated object (see Fig. \ref{f:bound}). It becomes a part of an endless 2D lattice of identical galaxies. The impact of this lattice of external galaxies can be controlled by adding a buffer of zeros around the galaxy being studied. However, adding a large buffer compounds the resolution problem, since for a fixed grid size ($N\times N$) the number of grid points that cover the galaxy are reduced.  Finding the right amount of buffering that will balance between these two opposing requirements is not a straightforward exercise. Even when the right amount of buffering is found, evaluating the impact of these constraints on lensing results is not trivial.\\

In this paper, we discuss an alternative method for performing the lensing calculations needed for ray-shooting. We focus on the methodology and postpone the applications to a forthcoming paper. 
This technique combines a smooth description of particles with well-defined lensing properties and a tree based domain decomposition. First we describe the principle of ray shooting through  FFT and the so-called 'Smooth Particle Lensing' (SPL hereafter) technique. Then, resolution effects and the capacities of SPL are investigated using realisations of analytic models such as softened isothermal spheres and ellipsoids.
Finally we conclude by discussing the intrinsic limitations induced by simulations regarding lensing predictions and present the future applications of the SPL technique.

\section{Ray Shooting}
To calculate the lensing properties of an object, we first adopt the thin lens approximation in which the mass distribution is collapsed to a plane perpendicular to the line of sight, the lens plane. Light is then assumed to travel along straight lines between the source, lens plane and the observer. Rays experience a discrete deflection at the lensing plane. The deflection angle can be calculated directly from the mass distribution.   This is an excellent approximation for galaxy and galaxy cluster size lenses.

\subsection{Lensing Formalism}
The deflection angle of a lens can be calculated from deflection potential, $\psi(\vec{\theta})$, of the mass distribution.
\begin{equation}
\label{eq:def_ang}
\vec{\alpha}(\vec{\theta})  = \nabla\psi(\vec{\theta}).
\end{equation}
This deflection potential is simply twice the 2D Newtonian surface potential and can be calculated from the convergence, $\kappa(\vec{\theta})$.
\begin{equation}
\label{eq:conv}
\nabla^2\psi(\vec{\theta}) =2\kappa(\vec{\theta}).
\end{equation}
The convergence is dimensionless quantity that can be calculated from the surface mass distribution on the lensing plane, $\Sigma(\vec{\theta})$, and geometric factors,
\begin{equation}
\label{eq:sig_c}
\kappa(\vec{\theta})= \frac{D_L D_{LS}}{D_S}\frac{4\pi G}{c^2}\Sigma(\vec{\theta}) = \frac{\Sigma(\vec{\theta})}{\Sigma_c} ,
\end{equation}
where $D_L$ is the angular diameter distance from the observer to the lens, $D_{LS}$ is the angular diameter distance from the lens to the source, and $D_S$ is the angular diameter distance between the observer and the source.  These factors are usually collected together into a term know as the critical density, $\Sigma_c=c^2D_{S}/4 \pi G D_L D_{LS}$.  When the surface density of a lens is greater than the critical density, ie $\kappa > 1$, a single source can have multiple images. Mapping a position on the lensing plane, $\vec\theta$, to the position of the source, $\vec\beta$, is done through the lens equation,
\begin{equation}
\vec\beta(\vec\theta)=\vec\theta - \vec\alpha(\vec\theta)
\end{equation}
The other properties of a gravitationally lensed system, magnification and shear, can be calculated using the distortion matrix, $A_{ij}\equiv \frac{\partial \beta^i}{\partial \theta^j}$.  Magnification is then $\mu(\vec{\theta}) = |A|^{-1}$, which is 
\begin{equation}
\label{eq:mag}
\mu(\vec{\theta}) = \frac{1}{(1-\kappa(\vec\theta))^2-|\gamma(\vec\theta)|^2}
\end{equation}
 and the two components of shear are defined as $\gamma_1=(A_{11}-A_{22})/2$ and $\gamma_2=A_{12}=A_{21}$. 

In the following sections, we first describe the classical method based on the Fourier transforms to compute the lensing potential of a simulated halos. Then, we introduce the 'Smooth Particle Lensing' technique.



\subsection{FFT}
Solving the Poison equation, (2), can be performed in Fourier space, where the equation becomes $\ell^2 \tilde{\psi}(\ell)=\tilde{\kappa}(\ell)$, where $\ell$ stands for the Fourier wavenumber.  This allows us to use Fast Fourier Transforms which use a grid based method for finding the Fourier transform of a field.  The first step is to put the particles onto a 2D grid.  Here we do this using a Cloud In Cell (CIC) routine.  At this stage the first undesirable effect of the FFT method is introduced. Placing the particles on a grid using CIC introduces a global smoothing with an effective CIC window function and limits the points where the lensing properties are calculated.  The CIC smoothing is a nuisance in strong lensing because this scale does not necessary correspond to scale that is significant to the investigation and so makes the interpretation of the results more difficult.  It should also be noted that using the FFT method for ray shooting is also widely used in weak lensing and cosmic shear simulations.  In these regimes the CIC smoothing is also problematic.  When considering the weak lensing regime one is typically interested the auto-correlations of the lensing field.  For such an analysis great care must be taken since it is widely known that using a simple FFT method to solve the Poisson can result in a loss of power on small scales due the CIC smoothing.  This loss of power affects a large range scales since the CIC window function extended in Fourier space.  This effect is well know and can be correct for \citep{2003MNRAS.341.1311S} , however it is not clear in the literature if this correction is applied.  In the strong lensing regime where the correlation function is usually not under investigation this loss of power may be less problematic, but the global smoothing is still, never the less, undesirable.  More problematic in the strong lensing case is the fact the lensing calculations are performed on a grid.

When studying the strong lensing properties of an isolated halo we are particularly interested in the  properties of the mass distribution in the very central regions of the halo since the impact parameters for light-rays are typically small.  At the same time, it is important to account for the distribution of mass outside this central region since it will also have an impact on the lensing signal.  This means that the entire galaxy must be placed onto a grid and the lensing properties calculated at each grid point.  However, however while locating and analysing images only only the central $\sim 0.1\% $ of the grid points are ever used.  This means that although the FFT is the fastest known method for computing the lensing properties at the N grid points, the fact that most of these points are not used becomes a significant loss of efficiency.  Restricting calculation results to grid points is also undesirable, since in order to study multiply imaged systems interpolation between the grid points can not be avoided since it is highly unlikely that all all images will sit on grid points.  On small scales, idea lensing solver would therefore be able to evaluate the lensing properties of a mass distribution at an arbitrary point.  For N-Body simulations the underlying density distribution is represented by point masses.  Without any smoothing, the shot noise from these point sources would dominate over the signals we are studying.  Therefore, although a one size fits all CIC smoothing is not desirable, a flexible smoothing is central to a lensing solving method.  There are two relevant smoothing scales. The first is the smoothing scale required to reduce shot-noise, which we expect to be linked to the local number density of particles.  The second smoothing scale is one that corresponds to the force resolution of the original N-Body simulation.  This is because density variation on scales smaller than this cannot be trusted to be of real significance.

On large scales the FFT method also suffers draw backs.  The introduction of repeating boundary condition has already been discussed.  The adverse effect of this for strong lensing studies is that a buffer needs to be placed around the galaxy halo.  Since the resolution limit is set by the maximum array that can be manipulated by a CPU, loading this array with zeros has a detrimental effect on resolution.  Once again, the impact of introducing repeating boundary conditions on weak lensing and cosmic shear needs to be considered carefully since this will impact correlations on large scales.  It is possible to construct a grid based method that does not introduce repeating boundary condition, however it complicates the calculation and these methods are not implemented in the current generation of ray shooting routines. 




\subsection{Smooth Particle Lensing (SPL)}

\begin{figure} 
\centering
\resizebox{0.95\columnwidth}{0.95\columnwidth}{\includegraphics{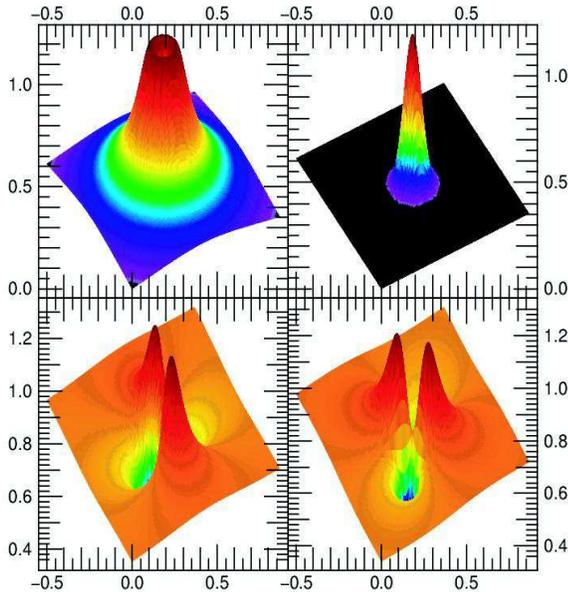}}
\caption{The properties of each of the particles in the simulation.  Each particle had a 2D Gaussian surface density distribution.  Top left shows the 2D gravitational deflection angle ($\alpha$), top right shows the convergence ($\kappa$) and bottom left and right show the two components of shear ($\gamma_1$ and $\gamma_2$).} 
\label{f:cloud} 
\end{figure}

Not all lensing calculations in numerical simulations use Fourier transforms, for example the method used by \cite{1998ApJ...494...29W} involves a tree based method to calculate deflection angles.
Here we suggest to go beyond such a calculation in order to compute all the lens quantities while  using an adaptive description of particles.

The `Smooth Particle Lensing'  technique is based on the on the discrete description of haloes in N-Body simulation and relies on the fact that each particle will contribute to the projected potential measured at a given point.  If $i$ labels the $i$-th particle of a simulated halo, the 2D deflection potential measured at $\bf r$ is given by:
\begin{equation}
\phi(\bf r)=\sum_i \phi_i(\bf r),
\label{e:sum1}
\end{equation}
where $ \phi_i(\bf r)$ is the potential created by a single particle.  Any linear function of the potential can also be expressed as a sum over particles.  The deflection angle is written as:
\begin{equation}
{\bf \alpha}(\bf r)=\nabla \phi(\bf r)=\sum_i \nabla \phi_i(\bf r)=\sum_i  {\bf \alpha}_i(\bf r),
\label{e:sum2}
\end{equation}
where $\bf \alpha_i(\bf_r)$ is the 2D deflection angles produced by a single particle. Likewise the convergence $\kappa$ and the shear $\bf \gamma$ are
\begin{equation}
\kappa(\bf r)=\sum_i  \kappa_i(\bf r),
\label{e:sum3}
\end{equation}
and
\begin{equation}
{\bf \gamma}(\bf r)=\sum_i  {\bf \gamma}_i(\bf r).
\label{e:sum4}
\end{equation}
Therefore, it is possible to derive the lensing properties of an arbitrary distribution of points by summing the contribution of individual particles. However we are left with two difficulties at this stage. First, we have to find a convenient single-particle lens model, with a 'clean' behaviour from both a physical and numerical  point of view. Second, we have to find an efficient way to perform the sums described in eqs. \ref{e:sum1}, \ref{e:sum2}, \ref{e:sum3} and \ref{e:sum4}. These two difficulties are assessed in the next two section.

\subsubsection{Smoothed Particles}
A natural way to describe the particles would be to describe their density as Dirac's $\delta(\bf r)$ functions, leading to 2D potential $\phi({\bf r})\sim \log(r)$ and deflection angles  $|\alpha ({\bf r})| \sim 1/r $. Not only such a choice would induce a noisy projected density but it creates singularities in the deflection angles, when close-encounters between a ray and a particle occur. Another choice would be to smooth the force applied to rays by the particles, e.g. by assuming $|\alpha ({\bf r})| \sim 1/\sqrt{r^2+\epsilon^2} $. It would correspond to the procedure applied in N-Body calculations. However, it still leads to singular convergence and shear for a single particle, making the method less powerful than it can be. Hence, we decided to start from a given functional form of a single particle density profile (i.e. its convergence $\kappa$), to construct its potential back by solving the Poisson equation, leading to expressions for the single-particle deflection angles and shear.

 Let us consider a particle with a mass $m_p$ and a lens configuration such as the deflector's redshift is $z_L$ and the corresponding critical density is given by $\Sigma_c$. Positions on the lens plane are given in terms of $\textit{physical}$ radii $r$ while angles can be recovered by 
 \begin{equation}
 \theta=\frac{r}{D_L(z_L)},
 \end{equation}
 where $D_L(z_L)$ stands for the angular-diameter distance of the lens. We chose to model the single-particle projected density by  a 2D isotropic 'Gaussian' function (see also figure \ref{f:cloud}). The related convergence is given by
\begin{equation}
\kappa(r)=\frac{m_p}{2\pi\sigma^2\Sigma_c}\exp\left(-\frac{r^2}{2\sigma^2}\right),
\end{equation}
where $r$ stands for the distance between the ray and the particle. The corresponding potential is given by 
\begin{equation}
\phi(r)=\frac{m_p}{4\pi\Sigma_c}(\log(\frac{r^4}{4\sigma^4})-2\mathrm{Ei}(-\frac{r^2}{2\sigma^2})),
\end{equation}
where $\mathrm{Ei}(x)=-\int^{\infty}_{x} \exp(-x)/x \mathrm{d}x$. The deflection angle is given by
\begin{equation}
\alpha(r)=m_p\frac{e^{-\frac{r^2}{2\sigma^2}}-1}{\pi r\Sigma_c}
\end{equation}
One can see in figure \ref{f:cloud} that the deflection angles (or equivalently the force applied to the ray by the particle) has a 'hollow' behaviour because of the finite but non-nil extension of the particle: the force rises as the ray gets closer to the 'border' (typically $r\sim 3\sigma$) of the particle then it adopts a behaviour in $1/r$ as expected.  The associated shear created by a single particle is given by:
\begin{equation}
\gamma_1(x,y)=\kappa(r)\frac{\left(x^2-y^2\right)
  }{ r^4 }\Gamma(r,\sigma)
\end{equation}
and 
\begin{equation}
\gamma_2(x,y)=2\kappa(r)\frac{x y }{r^4}\Gamma(r,\sigma),
\end{equation}
where
\begin{equation}
\Gamma(r,\sigma)=  r^2+2\left(1-e^{\frac{r^2}{2 \sigma ^2}}\right) \sigma
   ^2.
\end{equation}
The $\sigma$ parameter acts as a smoothing parameter and as $\sigma \rightarrow 0$, the single particle density profile tends to a Dirac-$\delta(r)$ function. This smoothing parameter can be a simple constant over all the particles, or in a more sophisticated way a function of the considered particle or ray. We show in section \ref{s:comp} how this choice can affect the final results.

\subsubsection{Tree 2D}
The single-particle model being set, an efficient way to perform the sums over all the particles' contributions remains to be found. Evidently it cannot be performed by direct summation, since the CPU consumption would scale as $N_\mathrm{rays}\times N_\mathrm{part}$. Furthermore an improvement in the simulation resolution, i.e. in $N_\mathrm{part}$, would strongly affect the computation time, while we want the simulation's resolution to be an advantage and not a limitation. For these reasons, we suggest to use a summation technique based on a tree based domain decomposition. These techniques have been widely used in numerical simulations to compute the 3D force created by an arbitrary distribution of particles (e.g. \citet{1986Natur.324..446B}, \citet{1988ApJS...68..521B}, \citet{dikaiakos}, \citet{1998ARAA..36..599B}). They are know to scale in a logarithmic way, i.e. they scale as $N_\mathrm{rays}\times \log N_\mathrm{part}$ in the current case. Here we apply the 2D version of this algorithm to compute all the lensing quantities created by  a set of particles distributed within the lens plane. Most of the following has been strongly inspired by the description made by \citet{dikaiakos} for the N-Body code PKDGrav.

The particles' distribution are organised following a 2D version of KD-Tree structures (see e.g. \citet{moore-tutorial}) or 2DTree. First, the space is divided in two subregions that contain the same number of particles. The whole box (or 'the root') is connected to two 'branches', which correspond to these two subregions. The same 'space-splitting' procedure is then applied to these smaller regions (or 'cells'), adding a new level to the tree with four new branches. In the end, the recursive application of this procedure leads to a binary tree, with final branches (or 'leaves') containing  $N_\mathrm{leaves}$ particles. We chose $N_\mathrm{leaves}=8$. While building the tree, each cell is being assigned an opening radius given by 
\begin{equation}
r_\mathrm{open}=\frac{2 r_\mathrm{com}}{\theta\sqrt{3}}+r_\mathrm{center}.
\label{e:opening}
\end{equation} 
The quantities $r_\mathrm{com}$ and $r_\mathrm{center}$ stand for the maximal possible distance between a particle inside the cell and respectively its center of mass and its geometrical center. The free parameter $\theta$ controls the opening radius, as explained below.

The lensing computation is performed by walking down the tree. We discuss this walking procedure for the computation of the force felt by a ray (or equivalently the deflection angles $\bf \alpha$) but it translates to the $\kappa$ and $\gamma$ computation. Let us consider the force felt by a ray at the point $\bf P$ on the lens plane. A given cell is opened if the distance $r_{\bf P}$ between its center of mass and $\bf P$ satisfies $r_{\bf P} < r_\mathrm{open}$. If the criterion is satisfied, the same test is performed on the two branches connected to the current cell. This recursion can be stopped in two manners.  First, the current cell is a leaf~: the force felt by $\bf P$ is obtained by adding the forces created by each particles of this cell. Second possibility,  $r_{\bf P} >r_\mathrm{open}$~: the force felt by $\bf P$ is obtained by computing the force applied by the whole cell, assuming a monopole with a mass equal to the number of particles inside the cell and centered on its center of mass. In other words, the effect of distant cells is modelled as the effect of a macro-particle with the same properties as a single one but scaled to the correct mass. 

The parameter $\theta$ appears as  a performance control parameter. In Eq. \ref{e:opening},  a small $\theta$ implies that all the cells are likely to be opened. In this case, the force computation is close to the direct summation of the interactions created by all the particles, resulting in slow computations. Conversely, a large $\theta$ would speed up the procedure but would also lower the accuracy of the computation.

At this stage, we end up with two free parameters: $\theta$ which controls the overall summation performance and $\sigma$ which introduce a finite spatial resolution.  In the following, we choose $\theta=0.7$: it led to accurate results while ensuring a good performance of the summation. The influence of sigma is more complex to establish and at the heart of the lensing computation at high resolution. 
\subsubsection{Adaptive Smoothing}
\label{s:adapt}
The final step to the SPL's setting is the implementation of an adaptive smoothing. It is clear that if particles sample the projected density, their 'extension' should be density-dependent. A constant smoothing length may result in over-smoothing in high density regions, while it would induce shooting noise in low density environment (see also sec. \ref{s:NSIS}). Hence, $\sigma$ should be a function of the local density of particles. Two strategies are possible, the first being that each dark matter particle is assigned a $\sigma$ by computing its local density. In practice, this involves the computation of the local density for $10^6-10^7$ particles, which can be highly CPU consuming.
 The second strategy consists of assigning the \textit{same} $\sigma$ to all the particles while its value depend on the local density of particles at the position of the light ray.  For different light-rays $\sigma$ changes.  This physically makes sense 
  since the density should be correctly estimated at the location of the light-rays, not at the dark matter particles location. Because the number of light rays is much smaller than the typical number of particles (at least an order of magnitude), this strategy significantly improves the computational efficiency.  It naturally implies that particles distant to the current ray have inappropriate $\sigma$. However, they only contribute to the deflection angle and, being distant, their influence varies as $m_p/r$ independently of $\sigma$, no loss of accuracy can be detected. We decide to adopt this second strategy based on the density at light-rays.
 
In our implementation, we chose to estimate this density by computing the distance $r_{\mathrm{sph}}$  to the $N_{\mathrm{sph}}$-th closest particle to the ray. In practice, this computation is made while walking the tree during the lensing computation. The density is simply estimated by :
\begin{equation}
\rho_{\mathrm{ray}}=\frac{N_{\mathrm{sph}}}{\pi r_{\mathrm{sph}}^2}.
\end{equation}
This procedure is equivalent to an SPH-density estimation with a top-hat Kernel. Even though more complex kernels may be used (see e.g. \citet{Li}), this simple choice is sufficient. Finally, we set a smoothing length for a given ray by applying the following relation:
\begin{equation}
\sigma=\sqrt{\frac{N_\sigma}{\pi\rho_{\mathrm{ray}}}}.
\end{equation}
The $N_\sigma$ parameter controls the smoothing strength and appears as a number of particle over which the smoothing is applied. Typically, we found that $N_\sigma\sim64-256$ give good results.

\subsubsection{SPL features}
Having described the technical aspects of SPL, we discuss briefly the interesting features of the method. 
First, a whole set of lensing quantities ($\psi$, $\bf \alpha$, $\kappa$ and $\gamma$) are directly computed from the particles' distribution and this can be extended to any linear function of the potential. Second, a key aspect of SPL is its flexibility in terms of geometry. No grid or any \textit{a priori} geometry is required for the sampling of the projected density. This methods takes fully advantage of the sampling performed by the simulation itself. At any point on the lens plane, we simply sum up all the particles contribution to any lensing quantity. This eliminates the use of interpolation procedures, which do not provide additional information compared to a coarser sampling. Furthermore, the computing power can be focused on an arbitrary region with any geometry. The result is an insensitivity of SPL to periodic artefacts, preventing the loss of computational power and memory in zero-padded regions (see also Fig. \ref{f:bound}). For instance, caustics can be mapped back by computing the deflections angle only along the critical curves. Also image distortions can be investigate at very high resolution by focusing the SPL calculation around an image spotted at lower resolution.

These features arise from the decoupling in SPL between the lens sampling and the computation of the lens' effect on the light rays. In N-Body calculations, the relevant quantity is the force exerted by particles on particles. Therefore, it makes sense to compute the forces at points where the density is sampled. In gravitational lensing, the relevant calculation is the influence of particles on light-rays. Hence, there is no reason to map the lensing quantities on a regular grid or at the particle positions. Grid-based methods do not achieve this decoupling easily, while it is natural for tree-based methods.  This makes tree-based (or more generally summation-based) methods much more appropriate to perform the computations, compared to N-Body calculations, where the pros and cons of grid-based and tree-based techniques can still be debated.

\section{Tests Using Analytic Distributions}
\label{s:comp}
\subsection{Non-Singular Isothermal Sphere}
\label{s:NSIS}
\subsubsection{Model}
To test the impacts finite resolution it is important to compare our measured lensing properties with analytic expectations.  For our initial test we focus on radially symmetric case. We begin by defining a convergence field from a Non-SIngular Isothermal Sphere (NSIS), 
\begin{equation}
\kappa(\theta)=\frac{\kappa_o}{\sqrt{\theta^2+\theta_o^2}}.
\end{equation}
Through out this paper we scale this distribution to be consistent with a halo of mass, $M_h=1\times10^{13} \rm M_\odot$, within a radius, $r_h=0.5 \rm Mpc$.  The halo is then assumed to be at a red-shift of z=0.8 and the background sources are assumed to be at a red-shift of z=3.0.

Solving the Poisson equation (\ref{eq:conv}) for the convergence field described earlier leads to the lensing potential,
\begin{equation}
\psi(\theta)=2\theta_o\kappa_o\Bigg(\sqrt{1+\frac{\theta^2}{\theta_o^2}}-\log\bigg(1+\sqrt{1+\frac{\theta^2}{\theta_o^2}}\bigg)\Bigg).
\end{equation}
and a deflection field of,
\begin{equation}
\alpha(\theta)=2\kappa_o\frac{\theta_o}{\theta}\bigg(\sqrt{1+\frac{\theta^2}{\theta_o^2}} -1\bigg).
\end{equation}
The subsequent shear and magnification are simply calculated from this deflection angle by constructing the distortion matrix, $\mathbf{A_{ij}}$ and using the definition above equation \ref{eq:mag}.

In order to draw conclusions about current numerical simulation we chose to study a NSIS were the core radius is 2000 times smaller than a typical virial radius for a galaxy of mass M.  In our case the M is $10^{13} \rm M_\odot $.  This core radius was chosen because it is approximately the smoothing scale expected in current N-body simulations.  These simulations contain roughly $10^7$ particles.  To accurately simulate an NSIS of this type under these conditions, we first converted the density distribution into a probability distribution function (PDF).  We then randomly sampled this distribution  $10^7$ times to obtain positions for the particles.  Finally by setting the mass of each of these points to $10^6 M_\odot$ we were able to produce realisation of our NSIS with particle sampling similar to that of an N-body halo.

\subsubsection{Radial Profiles}
 As an illustration, we begin by testing the impact of smoothing scale with a \textit{constant} smoothing scale $\sigma$. We do this by calculating the inverse magnification for a number of rays which are placed on concentric rings around the centre of the NSIS.  Such a test is crucial since magnification is the quantity that is most sensitive to variations in density, because it is a combination of second order derivatives of the lensing potential.  Furthermore, magnification serves to spot the location of critical lines and caustics. The radii of the rings are chosen so that they are distributed evenly in $\log(\theta)$ which allows us to cover a wide range of scales and hence take full advantage of the sampling flexibility of the SPL method. 

\begin{figure} 
\centering
\resizebox{0.95\columnwidth}{0.95\columnwidth}{\includegraphics{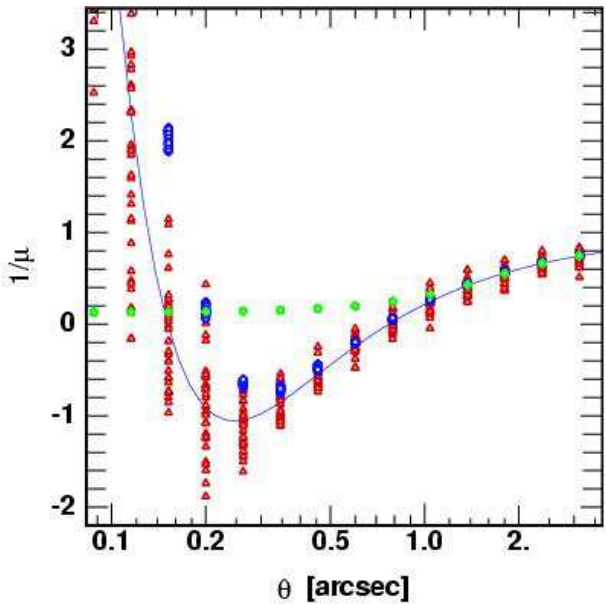}}
\caption{  The inverse magnification of a NSIS with a core radius of 0.037 arcsec ($r_{h}/2000$).  The solid line shows the analytic solutions.  The points are the results of SPL calculations on a distribution with $10^7$ particles. In each case a constant smoothing scale is used, the triangles (red) are for a smoothing of $\epsilon=0.0074$ arcsec, the diamonds (blue) are for $\epsilon=0.074$ arcsec and the circles (green) are for $\epsilon=0.74$ arcsec.}
\label{f:invmag}
\end{figure}

Figure \ref{f:invmag} shows the inverse magnification results.    For each case the smoothing was fixed to the same value over all the points calculated.  The green circles, blue diamond and red triangles are, respectively, for smoothing scales of 100, 1000, and 1000 times smaller than the galaxy radius.  From this we clearly see that a variable smoothing scale would be desirable.  In the outer roughly 2 arcseconds, we see that the green point accurately follow the analytic predictions with minimal spread, where as the red triangles suffer a substantial spread due to shot noise.  In the inner regions however the green points are incapable of finding the critical curves, points where the inverse magnification is zero, hence we would not recognise this as a system capable of having multiple images .  With the blue points we are able to accurately find the radial critical curve ($\sim 1$ arcsec), with very little spread, but fail to capture the tangential critical curve ($\sim 0.15$ arcsec).  In these very inner regions the red points do better but the spread is still substantial.

\begin{figure} 
\centering
\resizebox{0.95\columnwidth}{0.95\columnwidth}{\includegraphics{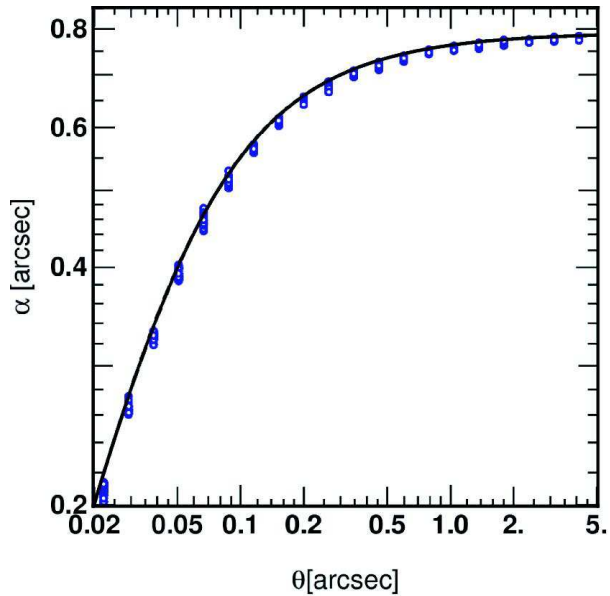}}
\caption{The deflection angle of a NSIS with a core radius of 0.037 arcsec ($r_{h}/2000$).  The solid black line is the analytic solution and the blue diamond symbols are those measured using the SPL method  with adaptive smoothing.}
\label{f:alpha} 
\end{figure}

\begin{figure} 
\centering
\resizebox{0.95\columnwidth}{0.95\columnwidth}{\includegraphics{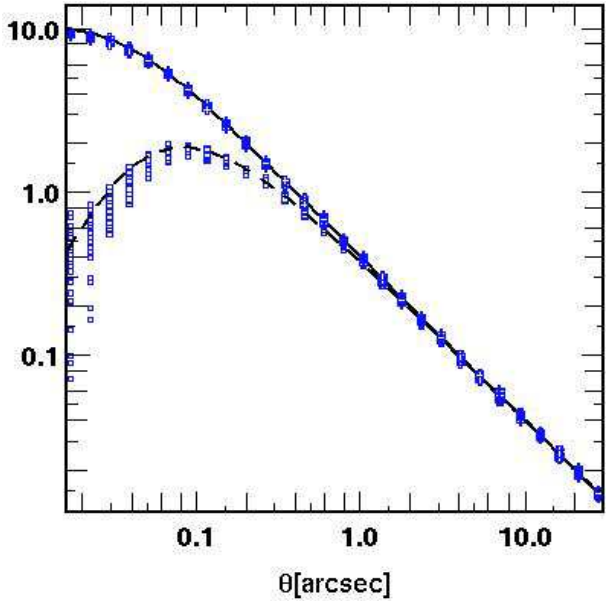}}
\caption{The solid black line shows the analytic solution for the convergence of the NSIS with a core radius of 0.037 arcsec ($r_{h}/2000$) and dashed line shows the magnitude of the shear ($|\gamma|$).  Overplotted are the convergence (blue squared) and the absolute shear (blue crosses) calculated using the SPL method on a distribution containing $10^7$ particles using adaptive smoothing ($N_\sigma=256$).}
\label{f:kap_gam}
\end{figure}

\begin{figure} 
\centering
\resizebox{0.95\columnwidth}{0.95\columnwidth}{\includegraphics{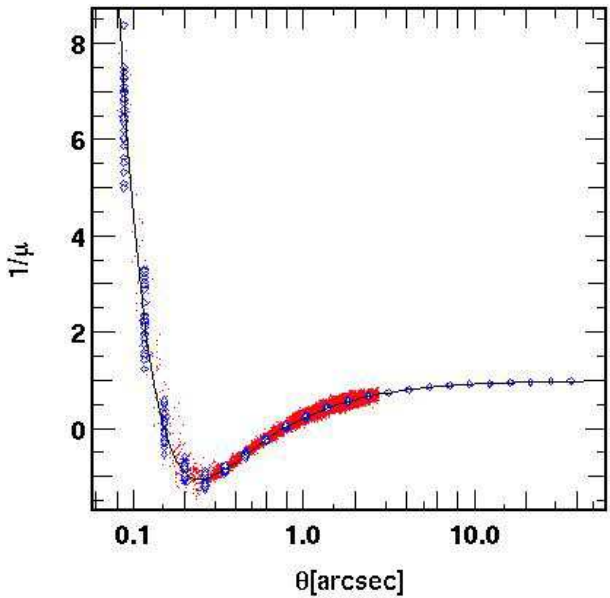}}
\caption{The inverse magnification of the NSIS with a radius of 0.037 arcsec ($r_{h}/2000$).  The solid black curve shows the analytic solution.  In red are the results of an FFT calculation performed on a $4096^2$ grid on a distribution made up of $10^7$ particles.  The blue diamonds show the SPL measurements on the same distribution using adaptive smoothing ($N_\sigma=256$). }
\label{f:invmag2} 
\end{figure}

From now on, we consider the case where an adaptive smoothing is set following the procedure described in section \ref{s:adapt}. Unless specified otherwise, we set $N_\sigma=256$.
Figure \ref{f:alpha} shows the deflection angle calculation for the same halo as in figure \ref{f:invmag}.  As expected we see that the deflection angle is easily calculated for a wide range of projected radii.  Figure \ref{f:kap_gam} we clearly see that the convergence and the shear fields are more sensitive to shot noise than the deflection angles, but they are still accurately reproduced.  Finally these are combined to calculate the inverse magnification shown in figure \ref{f:invmag2}.  Here both the results from the SPL method and the FFT method are shown.  We can see from this figure that the SPL does offers greater flexibility than the FFT method.  We are able to sample the  field at points of interest to us, in this particular case using concentric rings and we are able to control the spread thanks to the adaptive smoothing scale.

\begin{figure} 
\centering
\resizebox{0.95\columnwidth}{0.95\columnwidth}{\includegraphics{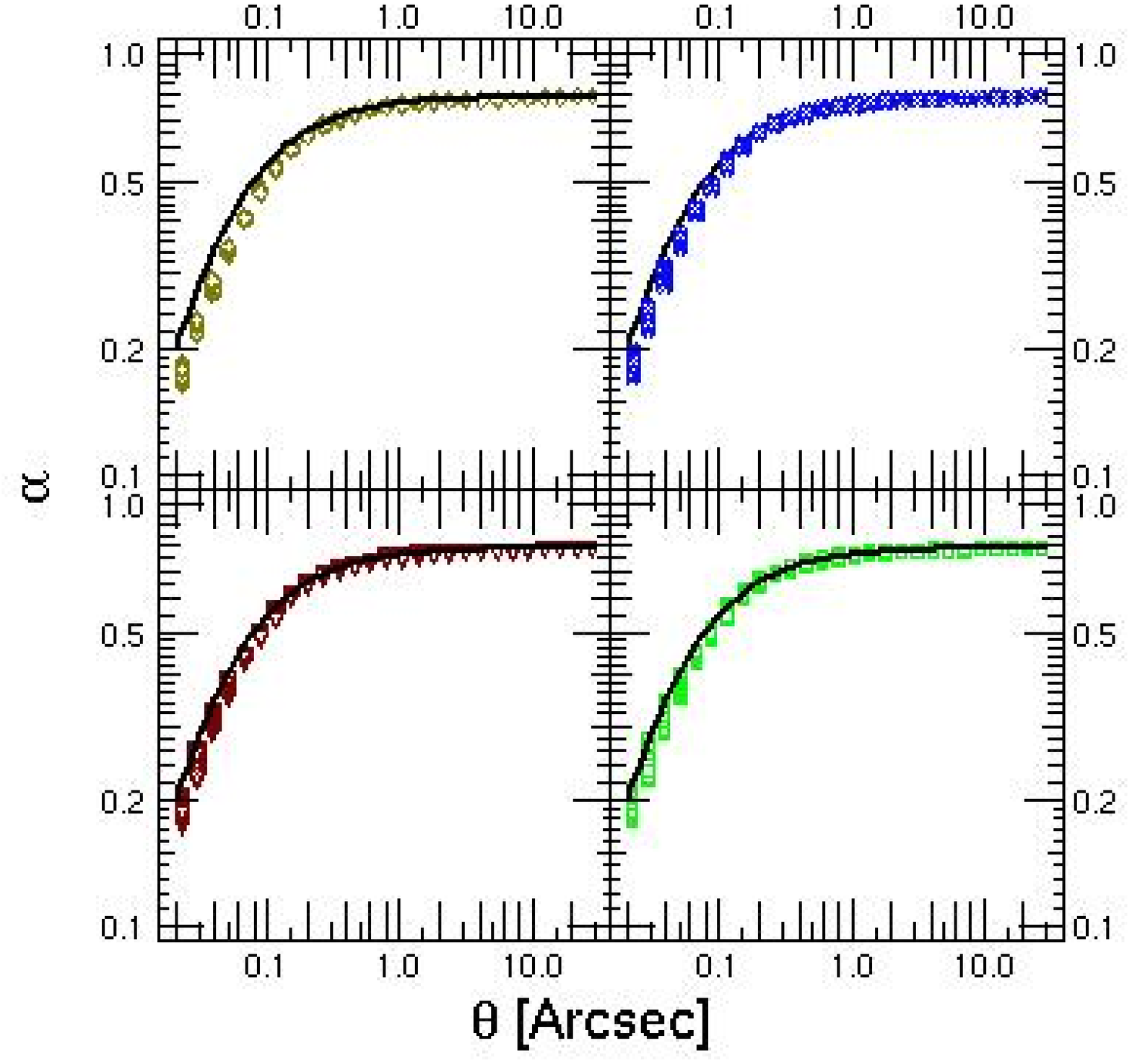}}
\resizebox{0.95\columnwidth}{0.95\columnwidth}{\includegraphics{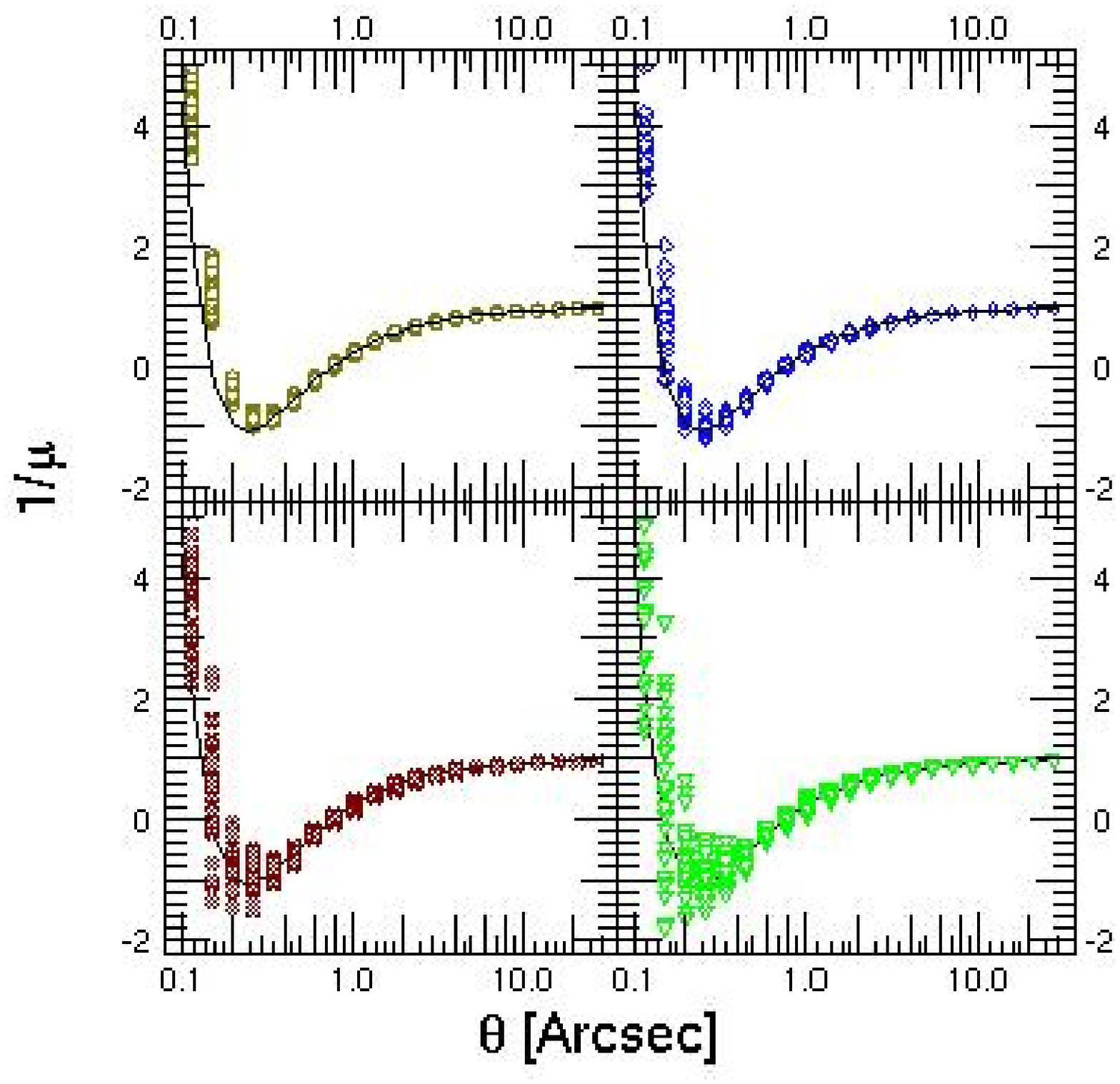}}
\caption{Deflection angle (\textit{Top}) and inverse magnification (\textit{Bottom}) of the NSIS with a core radius of 0.037 arcsec ($r_{h}/2000$). The model is sampled with $1e6$ particles. The SPL calculation is performed with $N_\sigma=256, 128, 64, 32$ (resp. gold, blue, red and green points). The solid lines stand for the analytic profiles.}
\label{f:million} 
\end{figure}

Finally, we show in figure \ref{f:million} the same calculations performed on the same model but sampled using only $10^6$ particles. We applied an adaptive smoothing with $N_\sigma=32,64,128,256$. The deflection angles are accurately computed, even though large smoothing values seem to slightly affect the calculation close to the halo's center. On the other hand, the inverse magnification is clearly more sensitive to the lower resolution of this halo, especially close to the center. Large smoothing value lead to a limited spread but the inner rise of the inverse magnification is incorrectly reproduced and occurs at larger radii than in the analytic solution. Conversely, small smoothing value seem to provide a better fit to the analytic magnification profile in these innermost regions, but the spread become very important. Consequently, constrains on the radial caustic would be highly uncertain even though they would be closer on average to the prediction. We conclude that such a halo is not suited to study the lensing properties of such models via SPL and we argue that it is very unlikely that another method would be able to compute correctly the lensing signal of the underlying model. Consequently, the limitation is clearly induced by the halo (or the N-Body simulation) and not by an inability to compute the signal properly.  

\subsubsection{Magnification Maps}

\begin{figure} 
\centering
\resizebox{\columnwidth}{\columnwidth}{\includegraphics{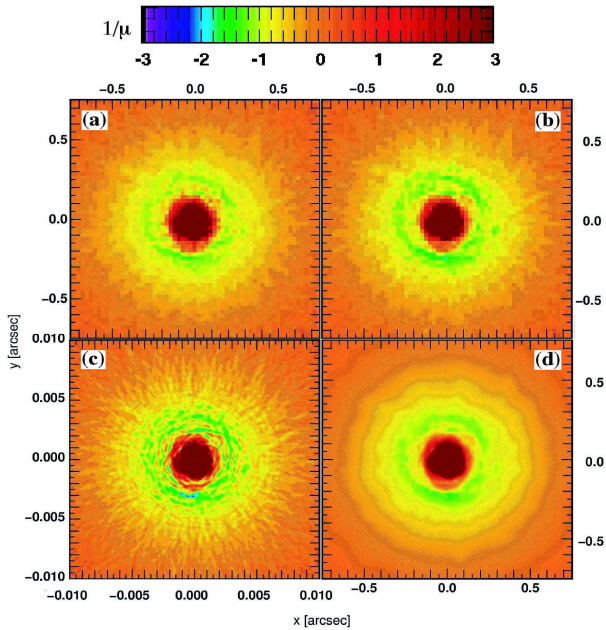}}
\caption{The inverse magnification for the NSIS.  Panel (a) shows the results from the FFT calculation on $4096^2$.  Panel (b) shows the result from the SPL calculation with a smoothing scale comparable to the CIC smoothing scale of the FFT calculation with sampling at the same grid points as the FFT calculation of.  Panel (c) shows the inverse magnification using the SPL routine where where the number of points where the lensing calculation is increased to $1024\times1024$  pixels and the smoothing scale is set to a constant $10^{-4}$. In this panel, coordinates are given in units  of the virial raidus. The same sampling is used in panel (d) as in panel (c), however in (d) an adaptive smoothing scale that is linked to underlying surface density is used.}
\label{fig:image} 
\end{figure}

The radial properties of the lensing fields discussed thus far give insight into the strengths is each method.  It is also very important to look at the lensing field in 2D. Figure \ref{fig:image} shows the inverse magnification for four difference cases.  Panel (a) shows the results of the highest resolution FFT grid calculation that we were able to perform ($4096\times4096$).  The pixel scale on the this zoom in is clearly visible since the image is made up of $\sim 160 \times160$ pixels.  Although it may be possible to optimise our routines further it is not likely that we will be able to increase the resolution by more than a factor of 2.  Panel (b) of figure \ref{fig:image} shows a set of SPL calculations performed at the same grid points as the FFT results of panel(a) and with a fixed smoothing scale that is comparable to the CIC smoothing.  These two 'low resolution' results are in good agreement with each other.  However, since the SPL calculations where only performed on the points shown, the SPL method was faster.  Panel (c) shows SPL results where the inverse magnification is sampled with $1024\times1024$ points on a polar grid  with a fixed smoothing scale of $10^{-4}$ in units of the virial radius.  Here the features from panels (a) and (b)  can been seen in detail and since they are persistent throughout all the calculations we must conclude that the 'flame' patterns are indeed real features unique to this realisation of the halo.  
They are there a result of shot noise due to finite mass resolution. Panel (d) shows the image of the inverse magnification when using adaptive smoothing with $N_\sigma=256$.  We see that by doing this we have managed to greatly reduce the 'flame' features. although the pattern is not perfectly symmetric as would be the case with infinite mass resolution.

\subsubsection{Critical and Caustic Curves}
\begin{figure} 
\centering
\resizebox{0.95\columnwidth}{0.95\columnwidth}{\includegraphics{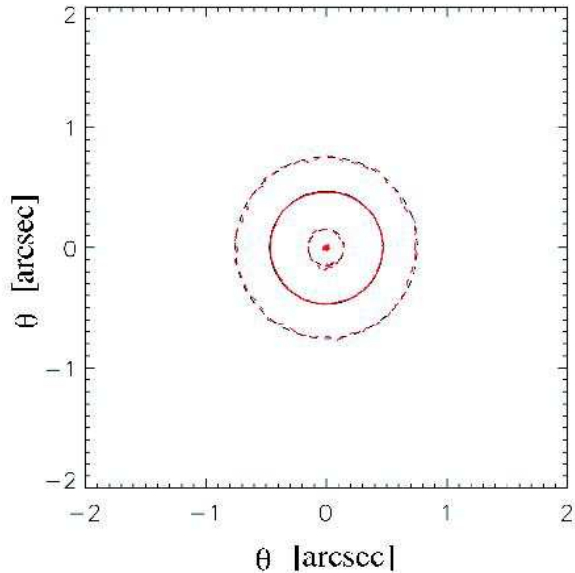}}
\caption{The critical and caustic curves for the NSIS.  The dashed red lines are the critical curves the solid red lines are the corresponding caustics.  The black curves are the analytic solutions and the red line are the results for SPL on a distribution with $10^7$ particles and core radius of 0.037 arcsec ($r_{h}/2000$).}
\label{fig:NSIS_crit} 
\end{figure}

The final remaining features of the NSIS are the critical and caustic curves.  These are shown in figure \ref{fig:NSIS_crit}.  Here the solid curves are the caustics and the dashed are the critical curves.  Shown in red are the curves we measure using our SPL method and in black are the analytic solutions for our NSIS. Clearly the two results agree very well and provide a 2D validation of the whole technique.
 
\begin{figure} 
\centering
\resizebox{\columnwidth}{\columnwidth}{\includegraphics{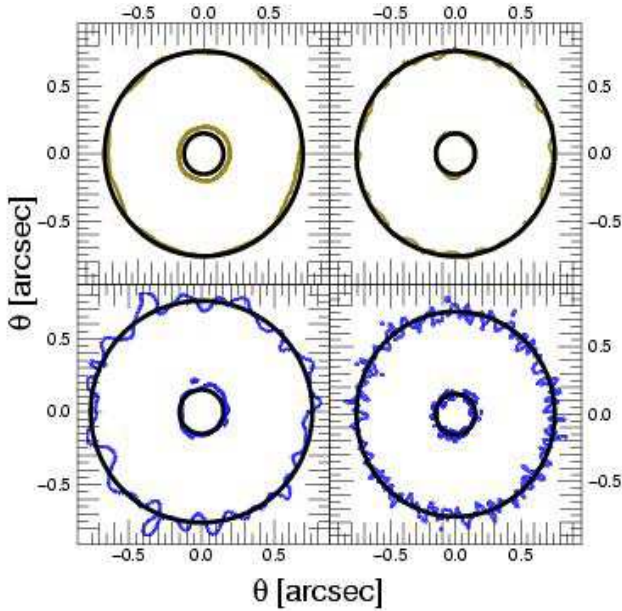}}
\caption{Critical curve locations for our template NSIS sampled with $10^6$ (left column) and $10^7$ (right column) particles. SPL calculation are performed with $N_\sigma=32, 256$ (resp. bottom and top row) on a 256x256 logarithmically sampled polar grid. The dashed lines stand for the analytically derived locations of the critical curves. The solid lines stand for the SPL calculations.}
\label{f:millioncrit} 
\end{figure}

The location of SPL's critical curves also permits us to investigate the halo's mass resolution issues. In Fig. \ref{f:millioncrit}, we show the critical lines computed for two haloes with $10^6$ and $10^7$ particles  and for two different smoothing ($N_\sigma=32,256$). Considering the $10^7$ halo first, increasing $N_\sigma$ removes the small scales features in the magnification map and reduces the spread around the analytic solution the critical curves. 
In general, the overall structure of these lines is well reproduced in all the cases. For the $10^6$ particle halo, the same behaviour occurs for the outer critical curve: more smoothing gives reduced spread and a better agreement with the analytic solution. However the inner curve drifts outwards from the correct location as $N_\sigma$ increases, even though the spread is reduced. Clearly, no good compromise can be found between resolution and noise control for a halo with this lower mass resolution.

We stress again that no grid-based interpolation has been used here to compute the deflection angle on the critical lines and map them back to caustics. A first computation has been performed to spot the critical lines (using the same procedure which results in panel (d) in Fig. \ref{fig:image} ). Then the exact deflection angles were computed along these critical lines only. 

\subsection{Non-Singular Isothermal Ellipsoid}

\begin{figure} 
\centering
\resizebox{0.95\columnwidth}{0.75\columnwidth}{\includegraphics{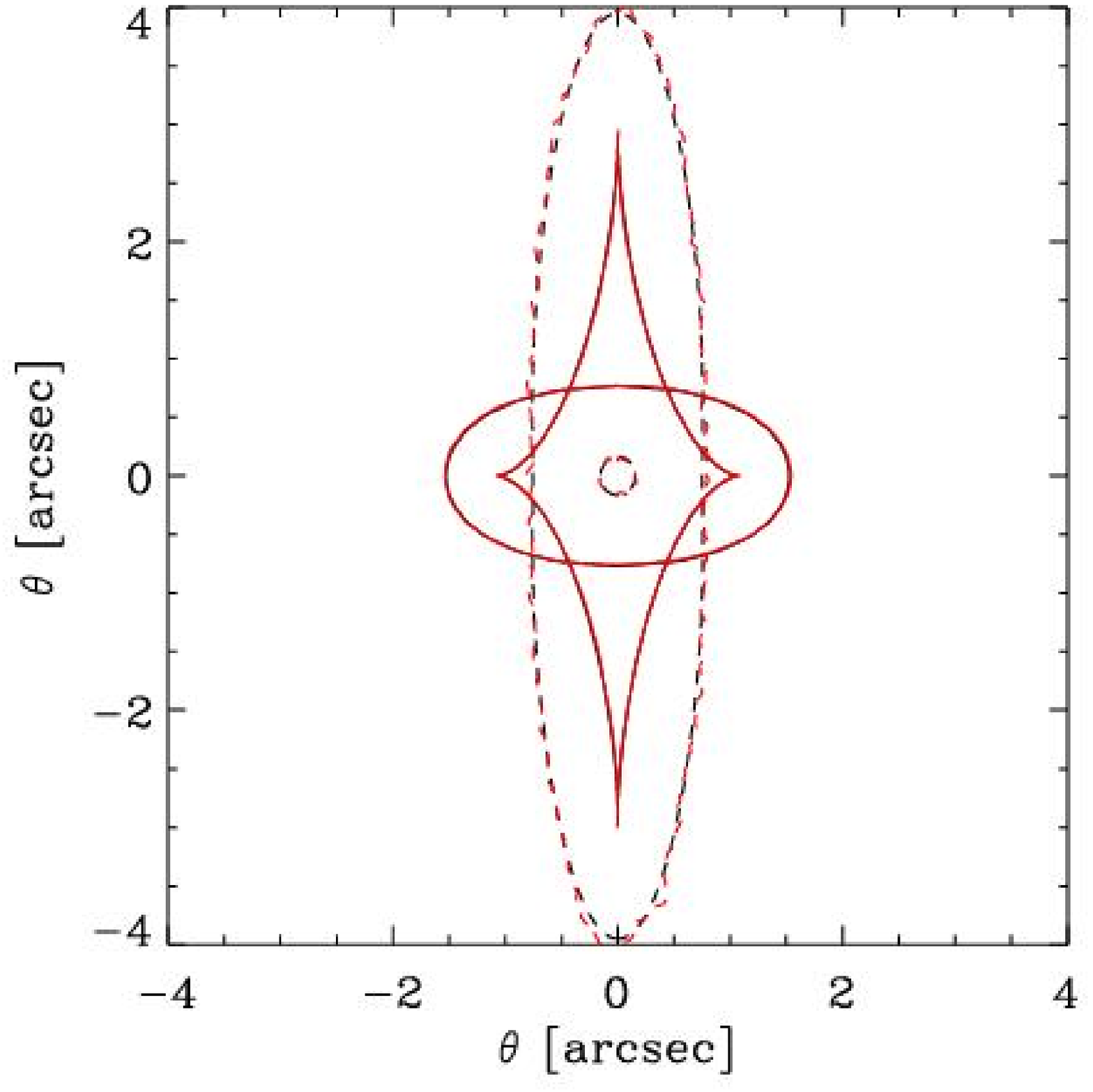}}
\resizebox{0.95\columnwidth}{0.75\columnwidth}{\includegraphics{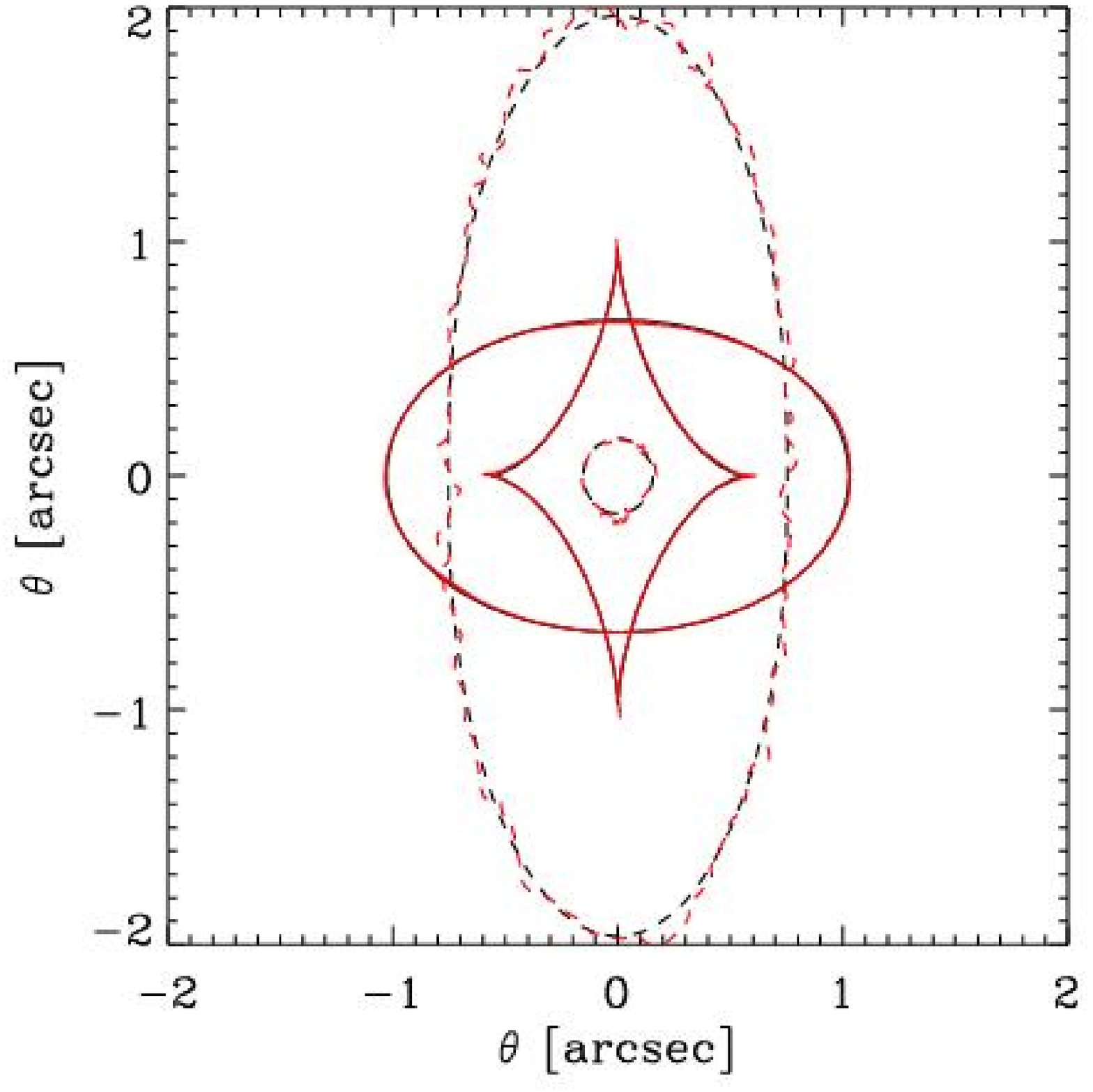}}
\resizebox{0.95\columnwidth}{0.75\columnwidth}{\includegraphics{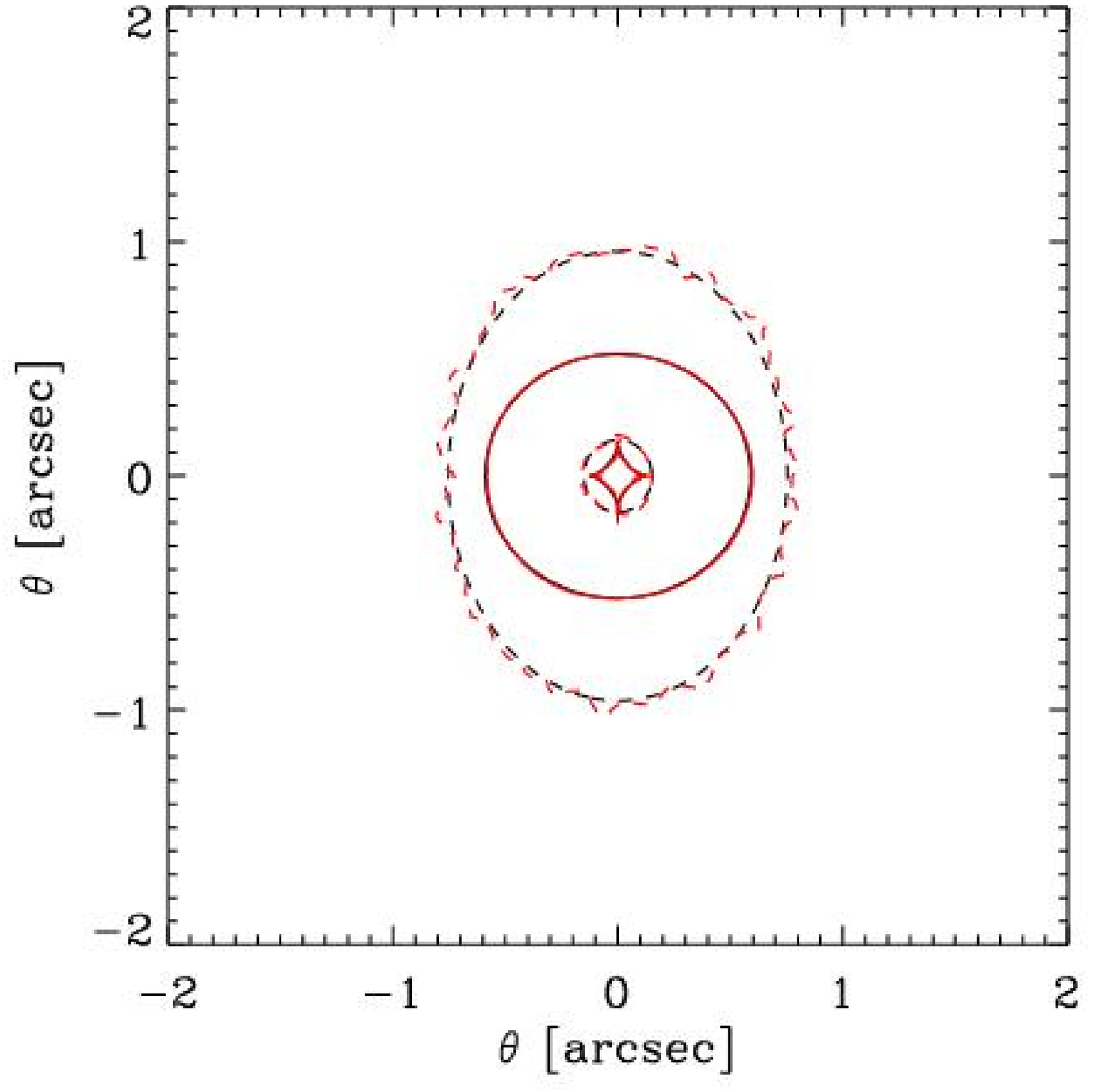}}
\caption{Critical (dashed) and caustic (plain) curve for the three different NSIE models: $e=0.2, 0.4, 0.8$ (from top to bottom). SPL calculations (red) are compared to analytic solutions (black).  NSIE models are sampled with $10^7$ particles and $M_h=10^{13} M_\odot$ and a core radius, $rc=r_{h}/2000$ with $r_{h}=500$ kpc. SPL calculations were performed with $N_\sigma=128$ on a 256x256 polar grid.}
\label{f:NSIE} 
\end{figure}

We continue our investigation of the SPL technique by breaking the spherical symmetry. We computed the magnification maps and the deflection angles for the Non-Singular Isothermal Ellipsoids (NSIE hereafter) described by \cite{1994A&A...284..285K}. These models are defined by the following projected density:
\begin{equation}
\kappa(\theta)=\frac{s_c\sqrt{e}}{2\sqrt{\theta_b^2+\theta_c^2}},
\end{equation}
where the scaling $s_c$ has been added to allow us to chose halo masses configurations. We have selected models with a halo of mass $M_h=1\times 10^{13} M_\odot$ with a main axis' length $r_h=0.5 $ Mpc. The lens configuration is the same as the one chosen for the NSIS, i.e. sources at $z=3$ and a lens at $z=.8$. Aside from the total mass and core radius, an additional parameter $e$,  the axis ratio, introduces ellipticity into the model through,
\begin{equation}
\theta_b=\sqrt{\theta_x^2+e^2\theta_y^2},
\end{equation}
where $\theta_x$ and $\theta_y$ are the direction in the $x-y$ plane.  We generated three $10^7$ particles' models with $e=0.2, 0.4.$ and $0.8$ and  Fig. \ref{f:NSIE} shows the critical and caustic curves location on the lens plane. SPL computations were performed with $N_\sigma=128$ on a 256x256 logarithmic polar grid with $10^{-4}<r/r_{h}<5\cdot10^{-2}$.  Because the spherical symmetry is broken, the radial caustic is not degenerate anymore: it allows to probe the accuracy of the outer critical curve's computation in a regime where the projected density is lower than in the central regions. Clearly the match between SPL calculations (shown in red) and the analytic solution (shown in black) is good and the lines are almost indistinguishable. Realistic dark matter haloes exhibits a certain level of triaxiality which should be easily handled with the current technique.  In particular, 'naked cusps' (i.e. cusps outside the radial caustics) are accurately reproduced, even for the most flattened models. 
Therefore image multiplicity statistics, the relative numbers of two-, four- images regions and naked cusp systems, should be well reproduced by the simulations.  

\section{Summary $\&$ Prospects}
We present a  method for computing the gravitational lensing induced by simulated gravitational lenses.  It solves the Poisson equation using a 2D-Tree decomposition technique combined with a description of simulation particles as 'extended clouds'. This so-called 'Smooth Particle Lensing' technique allows the \textit{direct} computation of any linear function of the deflection potential, such as the deflections angles, convergence, and shear. 
We test this method using analytic models of Non-singular isothermal spheres (NSIS) and ellipsoids (NSIE). The lensing properties are very accurately reproduced if the mass resolution is high enough. For instance, the caustic structure including naked cusps are correctly reproduced even for highly flattened systems and therefore accurate predictions of the image multiplicities are expected.

These tests showed that adaptive smoothing is \textit{necessary} if one wants to probe the whole lens plane to sufficient accuracy: a small smoothing scale does well in high density regions but induces a large scatter in low density environments. Conversely, over-extended particles smooth over  small-scale features such as the inner cusp and substructures.
In trying to achieve a compromise between smoothing and resolution
we found that the number of particles in simulated haloes can be critical to locating the critical and caustic curves. 
For instance, a NSIS with a 'core' as small as $r_{h}/2000$ and sampled with $10^6$ particles cannot be fully investigated in our fiducial lens configuration
($z_l=0.8$ and $z_s=3.$). This emphasises the need of high resolution haloes in order to accurately predict the strong lensing produced by these objects.

Because it combines noise-control with high resolution, this tool is particularly suited to studying the effect of substructures in dark matter halos, topics such as violations of the cusp caustic relation between image magnifications or predictions of image multiplicities (see e.g.\citet{astro-ph/0306238}, \citet{2006MNRAS.tmp..276A}). 
Because it does not suffer from periodic boundary conditions, SPL is naturally extended to weak lensing studies of isolated objects such as galaxy clusters.  In a large simulation box the method will adapt to put computational power where it is most needed, improving the accuracy of shear correlations over a large range of scales (from clusters to degree-scale cosmic shear surveys) while reducing CPU time.
Since a significant fraction of the N-body codes are based on tree structures, one could even imagine on-the-fly lensing calculations during the dynamical integration. 
A level of accuracy is achieved in our implementation of SPL where the calculation is effectively limited by the accuracy of the simulation used to find the mass distribution rather than the lensing code itself.

{\bf Acknowledgments}

{\sl  We   are  grateful   to E. Audit, R.Teyssier and A. Refregier for  useful comments
and  helpful  suggestions.  DA acknowledge support from the {\em HORIZON} project (www.projet-horizon.fr) and AA supported as part of the {\em DUNE (Dark Universe Explorer)}. }

\bibliography{SPL}
\bibliographystyle{mn2e}

\end{document}